\documentclass[final,11pt,3p,sort&compress]{elsarticle} %3p/5p, twocolumn
\makeatletter
\def\ps@pprintTitle{%
 \let\@oddhead\@empty
 \let\@evenhead\@empty
 \def\@oddfoot{\reset@font\hfil\thepage\hfil}%
 \let\@evenfoot\@oddfoot}

\biboptions{super}	%\citenum ohne super

\usepackage[T1]{fontenc} %umlaute in font encoding
\usepackage[latin2]{inputenc}
\usepackage{amssymb}
\usepackage{amsthm}

\usepackage{graphicx,cancel}
\usepackage{ulem}
\usepackage{color}
\usepackage{fixltx2e} %\textsuperscript{} and \textsubscript{}
\usepackage{microtype} %better spacing
\usepackage[
   separate-uncertainty  = true,
   mode = math,
   multi-part-units      = brackets,
]{siunitx} 
\usepackage{hyperref}
\hypersetup{
    colorlinks=true, % false: boxed links; true: colored links
    linkcolor=blue,  % color of internal links
    citecolor=blue,  % color of links to bibliography
}
\usepackage[all]{hypcap}  %link to the top of a figure

\renewcommand{\_}[1]{{}_{\mathrm{#1}}}

\begin{document}

\begin{frontmatter}

\title{Structural study of monolayer cobalt phthalocyanine adsorbed on graphite}
\author[ifw,tud]{M.~Scheffler}
%\ead{m.scheffler@ifw-dresden.de}
\author[tuc]{L.~Smykalla}
%\ead{\\lars.smykalla@physik.tu-chemnitz.de}
\author[ifw]{D.~Baumann}
\author[ifw]{R.~Schlegel}
\author[ifw]{T.~H\"{a}nke}
\author[tuc]{M.~Toader}
\author[ifw,tud]{B.~B\"{u}chner}
\author[tuc]{M.~Hietschold}
\author[ifw]{C.~Hess}
\address[ifw]{IFW Dresden, P.O. Box 270116, D-01171 Dresden, Germany}
\address[tud]{Institut f\"{u}r Festk\"{o}rperphysik, TU Dresden, D-01069 Dresden, Germany}
\address[tuc]{Chemnitz University of Technology, Institute of Physics, Solid Surfaces Analysis Group, D-09107 Chemnitz, Germany}

\date{\today}

\begin{abstract}

We present microscopic investigations on the two-dimensional arrangement of cobalt phthalocyanine molecules on a graphite (HOPG) substrate in the low coverage regime. The initial growth and ordering of molecular layers is revealed in high resolution scanning tunneling microscopy (STM). On low coverages single molecules orient mostly along one of the substrate lattice directions, while they form chains at slightly higher coverage. Structures with two different unit cells can be found from the first monolayer on. A theoretical model based on potential energy calculations is presented, which relates the two phases to the driving ordering forces.

\end{abstract}

%\begin{keyword}
%STM; cobalt phthalocyanine; organic molecule; epitaxy
%\end{keyword}

\end{frontmatter}

\section{Introduction}

Organic materials possess very intriguing properties, which make them outstanding from other functional materials used for spintronic applications, such as for example weak spin-orbit interaction and thereby implied slow spin relaxation times~\cite{Naber2007,Ding2010,Pramanik2007}. The interplay of single molecules on a substrate and their self-assembly properties were already investigated for various kinds of molecules and substrate materials~\cite{Forrest1997,Hooks2001,Smykalla2012,Walzer2001,Takada2004,Guo2010,Jiang2011,Hao2011,Toader2011,THPPstruc}. Their small size makes them very suitable candidates to serve as building blocks of functional nanostructures. Metallophthalocyanines are in the center of current research because of their magnetic and electronic properties. The spin induced by the metal ion conveys important properties for the application in spintronic devices, for example memory elements and logic gates~\cite{Prinz1999,Collier1999}. An important method for real-space topographic and spectroscopic measurements of molecular structures is the scanning tunneling microscopy, which images the electronic structure of a surface and its adsorbates with sub-\AA\  resolution. This method requests the molecules residing on a conducting substrate in order to allow a current flow through the molecule. However, if the substrate is metallic, it may substantially alter the molecular electronic properties~\cite{Hu2008,Li2010,Toader2010a,Zhang2011,Petraki2011}. A solution to this is to use only weakly conductive substrates, such as thin insulating layers on a metal substrate~\cite{Ruggiero2007} or intrinsically weakly conducting materials~\cite{Scarfato2008}, where graphite is a prime candidate.

In this study, we investigate cobalt-phthalocyanine (CoPc) molecules on a highly-ordered pyrolytic graphite (HOPG) substrate in the low coverage regime using scanning tunneling microscopy (STM). In the submonolayer regime, single molecules oriented along the substrate lattice directions as well as small chains of molecules can be found. At higher coverages, the molecules form typical ordered structures, which differ depending on the parameters of the sample preparation. Annealing of the sample leads to a higher mobility of the molecules on the surface and enhances the degree of order. The two dominant, highly ordered structures exhibit a long-range order. The grain boundaries between them form either a stripe-like pattern or two-dimensional ordered arrays. An epitaxy model combined with molecular mechanics calculations is presented and compared with the experimental results. A very good agreement between the measured data and simulated structures could be found.

\section{Methods}
\subsection{Experimental details}

CoPc molecules (Fluka, purity $> 97$\,\%) were degassed for purification and deposited on a HOPG substrate via organic molecular beam epitaxy in a Knudsen cell~\cite{Kowarik2008,Herman1988}. The evaporation took place under UHV conditions at evaporator temperatures between \SI{350}{\celsius} and \SI{400}{\celsius}. A quartz microbalance monitored the thickness of the deposited molecular film. The HOPG substrate was freshly cleaved and annealed up to \SI{1600}{\celsius} prior to the deposition. The substrate temperature during the evaporation process was either kept at room temperature or held constant in a temperature range up to \SI{100}{\celsius}. Also, further annealing of the arranged molecules at temperatures up to \SI{200}{\celsius} was performed. To guarantee a high reproducibility of the sample preparation and the results, the investigations took place in three different microscopes and the in situ-preparation was done in three UHV chambers with their corresponding preparation devices.

The molecules are very mobile on the substrate surface. The mobility is reduced for the molecules assembled within a molecular layer, but to achieve stable imaging it is necessary to freeze the molecular motion by low temperatures (below \SI{30}{\kelvin} in our experiments). Still, single molecules can be moved accidentally by the influence of the scanning tip [as can be seen at the separately lying molecules on the left side of the figure~\hyperref[fig:below]{\ref*{fig:below}(a)}]. The shape of these molecules is apparently strongly distorted and their position is ill-defined in the images. Therefore, throughout this paper we only discuss molecular arrangements where the overall single molecular shape is preserved in the images and where backward and forward scanning directions yield the same molecular arrangement. Significant accidental movements of the discussed molecule arrangements by the scanning tip are therefore irrelevant. Furthermore, we did not observe a significant change of topographic findings as a function of scanning parameters such as tunneling voltage/current or scan speed.

Two home-built microscopes and a commercial VT-Omicron system were used for investigation. The temperature during the measurements was \SI{5}{\kelvin}/\SI{20}{\kelvin}/\SI{30}{\kelvin} depending on the microscope, the pressure in the UHV systems was less than $\SI{1E-9}{mbar}$. All shown STM pictures were obtained with electrochemically etched tungsten tips. The indicated bias voltages were applied to the sample, so that the occupied states are imaged at negative voltages. STM images were processed with the WSxM software~\cite{Horcas2007}.

\subsection{Model calculations}

\begin{figure}[!tb]
\centering
\includegraphics[width=0.5\textwidth]{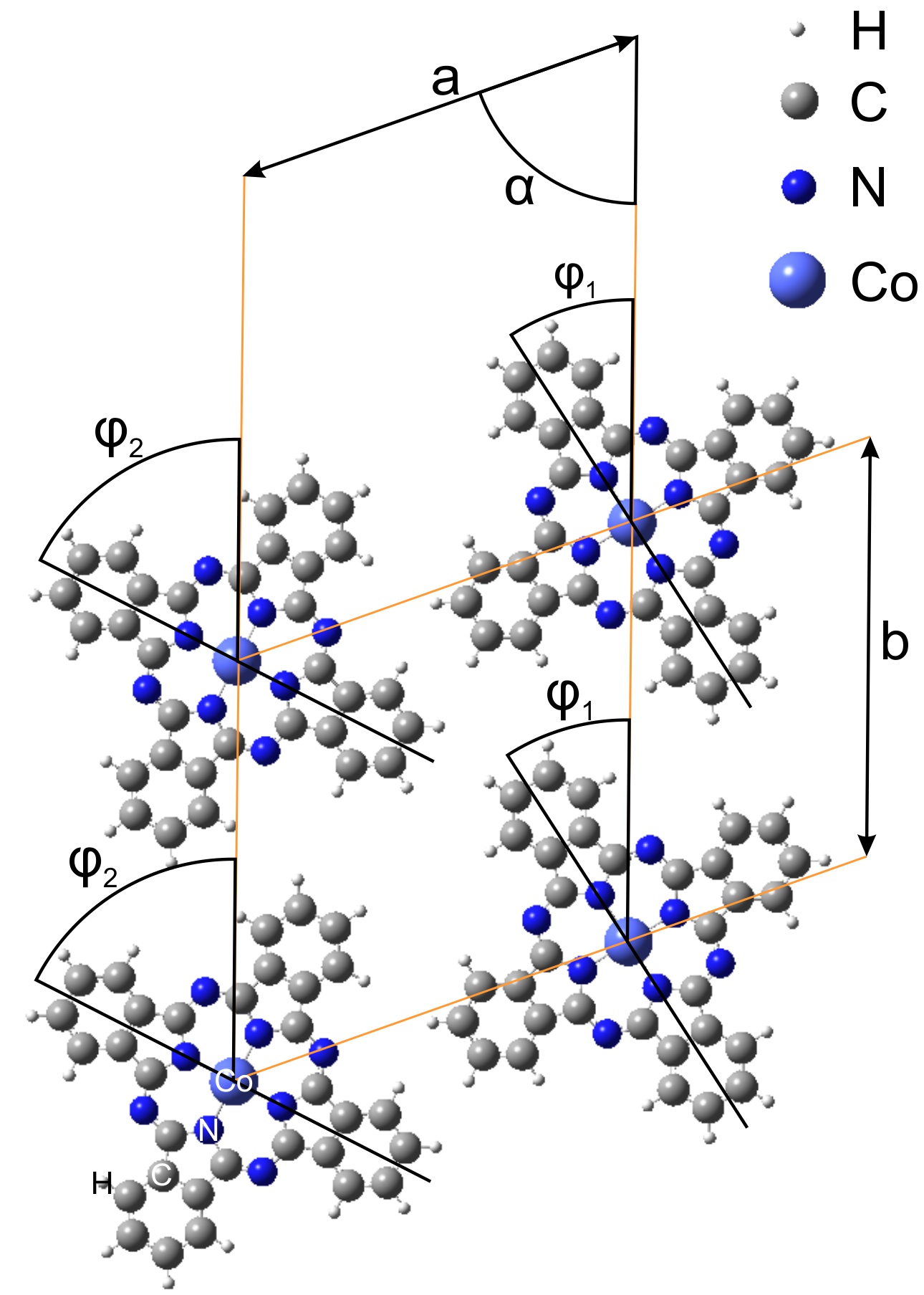}
\caption{One observed structure of CoPc molecules and assignment of the angles for UFF calculations and measured data.}
\label{fig:skizze}
\end{figure}

To gain insight in the delicate energetics which are responsible for the self-organized structure formation of the CoPc molecules, molecular mechanics simulations were applied. The calculations were carried out with the Gaussian 03 package~\cite{Frisch2004}. The molecular structure of the CoPc molecules was obtained by minimizing the energy using density functional theory. Potential energy surfaces for an arrangement of four CoPc molecules were calculated with the Universal Force Field (UFF)~\cite{Rappe1992}. Thereby, the calculated total energy is the sum of all non-covalent energies from the Pauli repulsion and van der Waals interactions, which are expressed in a 6-12-Lennard-Jones form, and the electrostatic interaction between each non-bonded atom. The binding energy is calculated by subtracting the total energy of the arrangement from the sum of the energies of the isolated parts. To model the molecular lattice, four CoPc molecules are placed on the corners of a parallelogram which describes the unit cell as shown in figure~\ref{fig:skizze}. This model is sufficient to give qualitative results for the intermolecular interactions for one molecule with its surrounding molecules in the molecular layer. No substrate is included in the calculations, and the plane of the molecules is restricted to lie in the plane of the layer. To simulate different possible structures, calculations were done where either all molecules have the same orientation or where the orientation of the molecular axes can differ between two neighboring molecules. To obtain potential energy surfaces two parameters of the arrangement were varied while the rest remained fixed.

\section{Results and Discussion}
\subsection{Experimental Data}

\begin{figure}[!tb]
\centering
\includegraphics[width=0.8\textwidth]{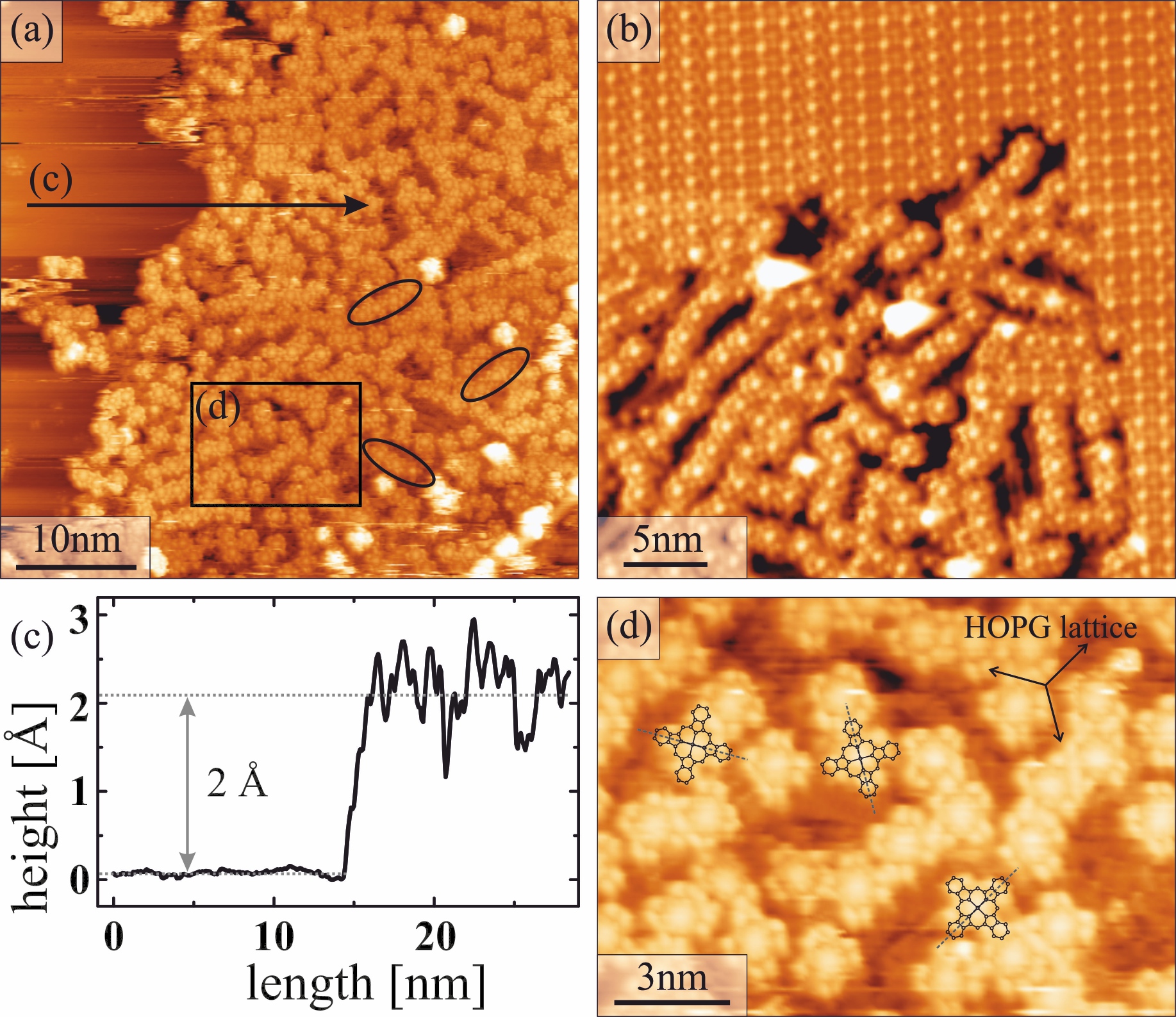}
\caption{CoPc molecules adsorbed on a HOPG substrate in a low coverage. They lie flat on the surface, as can be seen on their four-fold symmetric appearance. A tendency to form chains is observed.
(a) The STM image shows a large disordered agglomeration of individual CoPc molecules. Some of them form short chains of three or four molecules, as in the encircled areas. On the left side of the image the bare substrate is visible.
The sample was held at room temperature during deposition and not annealed afterwards. Tunneling conditions: $U = \SI{2.5}{V}$, {$I = \SI{0.07}{nA}$}, $T = \SI{21}{\kelvin}$ (b)  At the right and top of this image a highly ordered region can be seen [compare with Fig.~\hyperref[fig:phases1]{\ref*{fig:phases1}(a)}]. At its edge long chains of molecules exist. The substrate of this sample was {heated} up to \SI{100}{\celsius} during the evaporation process. Tunneling conditions:  $U = \SI{2.5}{V}$, $I = \SI{0.07}{nA}$, $T = \SI{21}{\kelvin}$ (c) Line profile along the arrow in (a) shows a height of the molecules of about \SI{0.2}{nm}.
(d) Zoom of rectangular area depicted in figure~\hyperref[fig:below]{\ref*{fig:below}(a)}. The lobe structure of the individual molecules is clearly resolved. Many molecules show an orientation as depicted.}
\label{fig:below}
\end{figure}

\begin{figure}[!tb]
\centering
\includegraphics[width=0.8\textwidth]{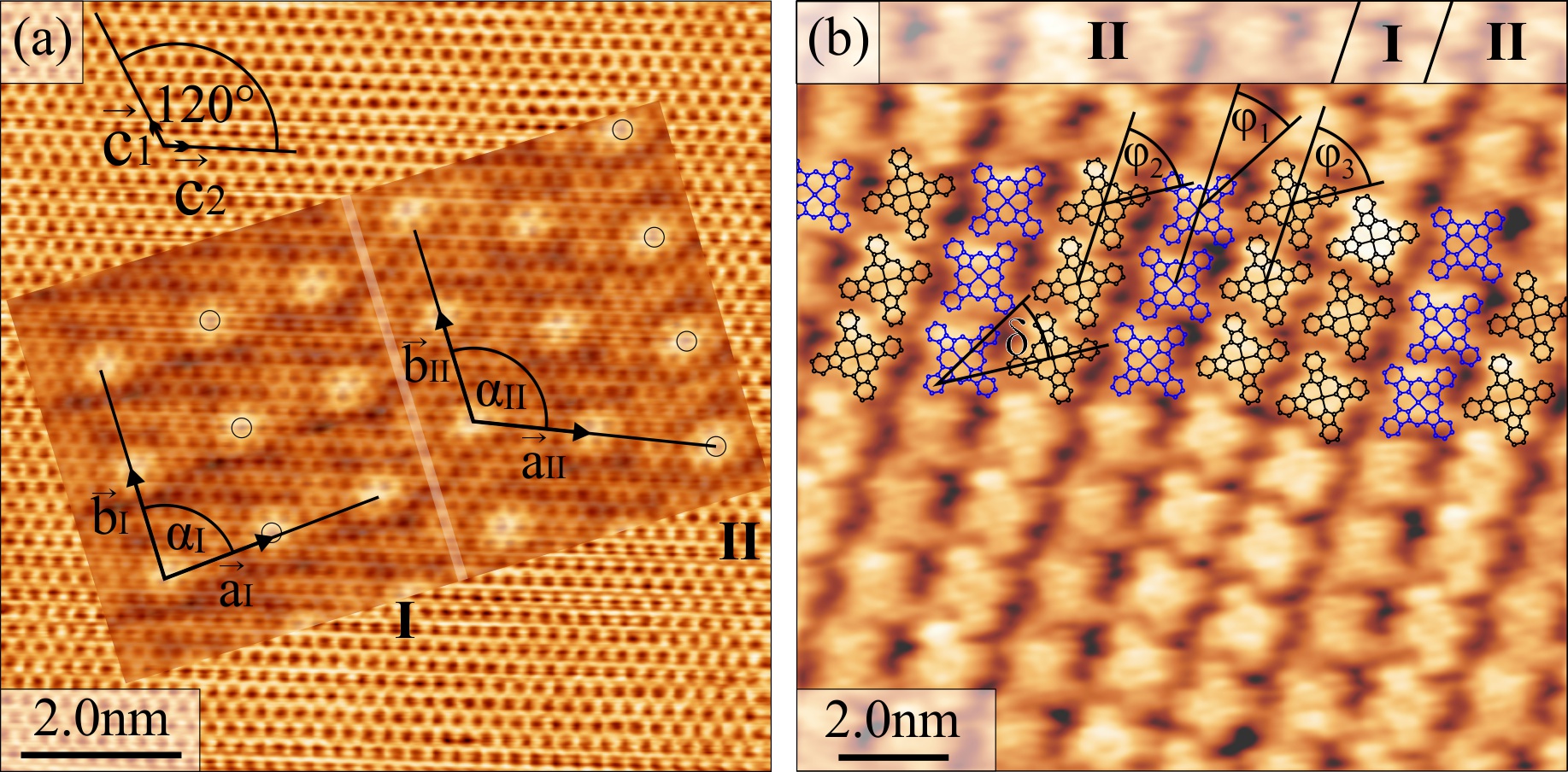}
\caption{(a) An overlay of both phases is compared with the HOPG substrate. The image of the substrate was taken separately on the same sample at a nearby uncovered area. The specific adsorption sites remain unknown, therefore only angles and lattice constants are compared here. They can be found in the text. Tunneling conditions: $U = \SI{2.5}{V}$, $I = \SI{0.07}{nA}$, $T = \SI{21}{\kelvin}$ (b) The rotation of the molecular axes with respect to each other differs depending on the phase. In phase~I all molecules are oriented the same way. In phase~II the angle between two neighboring molecules is $\delta = \SI{29}{\degree}$. This figure shows the 2nd ML after annealing at elevated temperature for a few hours, but the observation holds also true for the 1st ML. The extreme case is shown here where phase~I consists only of one unit cell. Tunneling conditions: $U = -\SI{1.5}{V}$, $I = \SI{0.1}{nA}$, $T = \SI{30}{\kelvin}$}
\label{fig:phases2}
\end{figure}

Figure~\ref{fig:below} shows representative areas of CoPc molecules deposited on HOPG. The coverage is less than one monolayer. Individual molecules appear in figure~\hyperref[fig:below]{\ref*{fig:below}(a)} in a flower-like shape with four lobes, as expected from their chemical structure. They do not exhibit a common orientation, but orientations along the plotted sketches as well as \SI{60}{\degree} rotated are preferred for most of the molecules [Fig.~\hyperref[fig:below]{\ref*{fig:below}(d)}]. The molecules have an apparent  height of about \SI{0.2}{nm}, as can be seen from the linescan [Fig.~\hyperref[fig:below]{\ref*{fig:below}(c)}] along the arrow in figure~\hyperref[fig:below]{\ref*{fig:below}(a)}. The symmetry of the molecules together with their height implies that they are adsorbed flat on the surface.
Some molecules form ensembles of two or more, as can be seen in the encircled areas in figure~\hyperref[fig:below]{\ref*{fig:below}(a)}. The density of molecules in the covered area is 0.26\,molecules/nm$^2$.
In the STM images, at positive bias voltage the center of each molecule, which is assigned to the metal atom, often appears brighter due to an increased density of empty states at the cobalt site of the molecule. At positive bias we expect a shape of the molecules accordant to the $\pi$ electron orbital of the ligands and a Co d$\_{z^2}$ orbital of the molecule, while at negative bias only the $\pi$ orbital of the ligands is imaged only [e.g. in Fig.~\hyperref[fig:phases2]{\ref*{fig:phases2}(b)}], similar to CoPc molecules in a stacked compound~\cite{Martinsen1985} or adsorbed on a gold substrate~\cite{Li2010,Lu1996,Zhao2005}. The influence of the HOPG substrate is expected to be very small due to the small binding energies of a layered material between its layers as well as towards an adsorbate on the top layer.

In order to enhance the coverage towards $\approx 1$ ML we prepared a second sample where we used a 25~K higher evaporator temperature (i.e. with a higher evaporation rate) for approximately the same time as has been used for the sample in figure~\hyperref[fig:below]{\ref*{fig:below}(a)}. Additionally, we heated the substrate to \SI{100}{\celsius} during the evaporation  in order to enhance the mobility and introduce a faster and better ordering of the molecules.
Under these conditions, the onset of an ordering of the molecules into a closed two-dimensional arrangement takes place as can be inferred from figure~\hyperref[fig:below]{\ref*{fig:below}(b)}. At this coverage still patches with a relatively lower coverage ($\approx 0.24$\,molecules/nm$^2$) are present at the borders of the close-packed areas. These areas contain chain-like linear arrangements of molecules. Apparently, this is an intermediate state between the disordered low-coverage and the ordered high-coverage state. The density of the molecules is the same as for the unordered agglomeration of molecules [compare with Fig.~\hyperref[fig:below]{\ref*{fig:below}(a)}]. In the close-packed areas the density is about 0.4\,molecules/nm$^2$, which is nearly twice as high as in the short-range ordered regions.

\begin{figure}[!tb]
\centering
\includegraphics[width=0.8\textwidth]{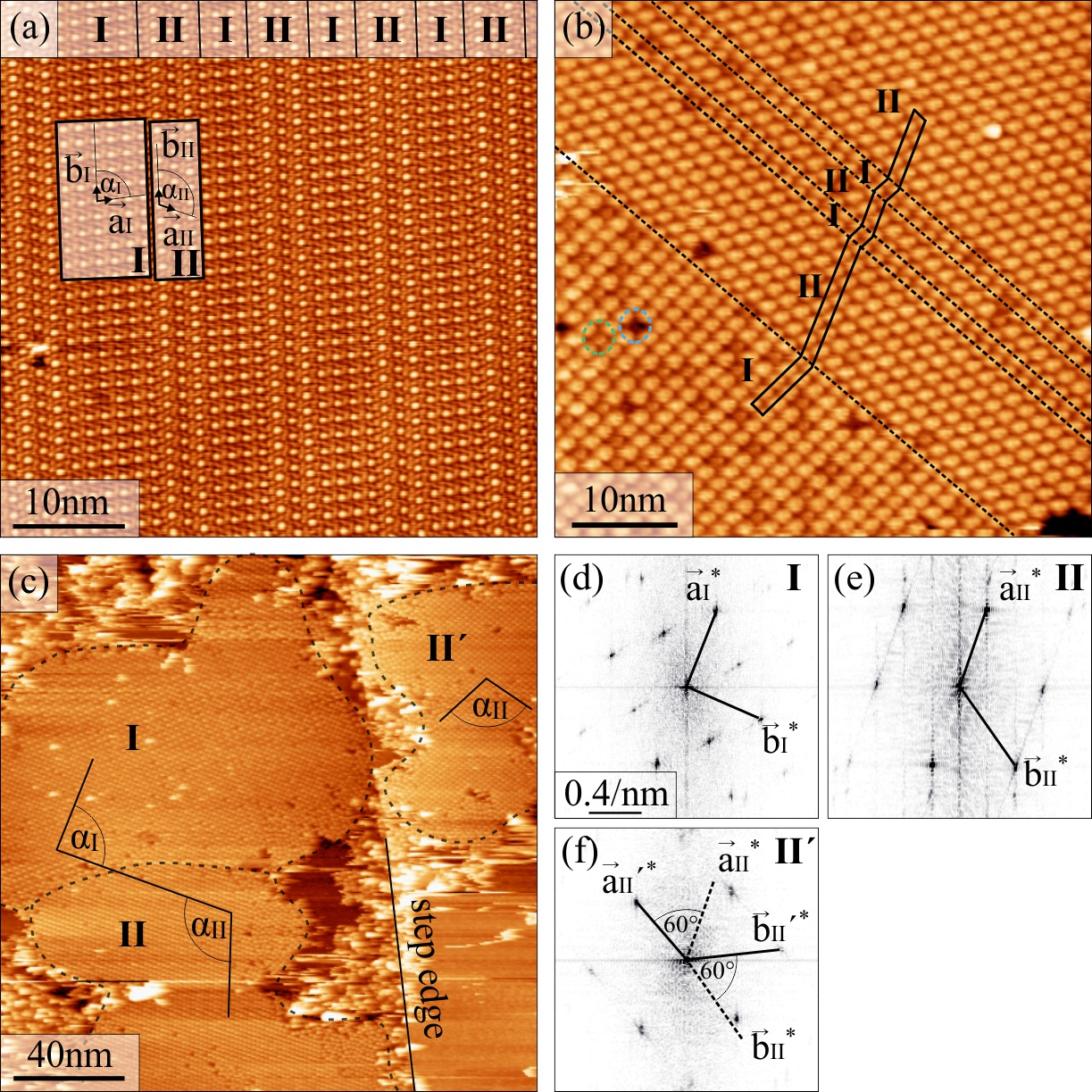}
\caption{(a) The area covered completely with molecules of the first ML shows a stripe-like pattern with two phases, which extend over long distances. The substrate of this sample was heated up to \SI{100}{\celsius} during the evaporation process. Both phases extend along a common direction. Perpendicular to this they alternate, whereupon the number of molecules in each phase is either even (I) or odd (II). Tunneling conditions: $U = \SI{2.5}{V}$, $I = \SI{0.07}{nA}$, $T = \SI{21}{\kelvin}$ (b) On another sample, which was post-annealed at elevated temperature for a few hours, large undisturbed regions of each phase are observed. Boundaries between phase I and II are marked with dashed lines. The image shows a second ML of CoPc. A missing molecule in the second layer is marked with a green circle and a small protrusion from the first layer molecule can be seen at this location, whereas holes through both layers appear darker (blue circle). Tunneling conditions: $U = \SI{1}{V}$, $I = \SI{0.08}{nA}$, $T  = \SI{30}{\kelvin}$ (c) The STM image shows areas of only one phase at larger scale. These regions are sometimes rotated with respect to each other by \SI{60}{\degree}. Tunneling conditions: $U = \SI{1.5}{V}$, $I = \SI{0.08}{nA}$, $T = \SI{30}{\kelvin}$ (d--f) FFT images of the regions depicted in (c). Region II and II´ are \SI{60}{\degree} rotated with respect to each other, as can be seen in a rotation of the lattice spots. The adjacent regions I and II exhibit a common lattice direction ($|\vec{a}\_I^*|=|\vec{a}\_{II}^*|$).}
\label{fig:phases1}
\end{figure}

In figure~\hyperref[fig:phases1]{\ref*{fig:phases1}(a)}, a highly ordered area with a large molecular layer of monolayer height is shown. The molecules appear here as spots of high local density of empty states in their center (originating from the cobalt atom d$\_{z^2}$ orbital), encircled by a fourfold symmetric pattern of the ligands. There are only few lattice defects in this area, such as missing molecules or single additional molecules adsorbed on top of the first layer. We found that two stripe-like domain types coexist in the first monolayer. We label them phase~I and phase~II, as assigned in figure~\hyperref[fig:phases1]{\ref*{fig:phases1}(a)}. They are independent of each other as they occur with the same lattice parameters in stripe patterns as observed in figure~\hyperref[fig:phases1]{\ref*{fig:phases1}(a)} as well as in large areas of just one domain [Fig.~\hyperref[fig:phases1]{\ref*{fig:phases1}(c)}]. They do not show a mirror symmetry on the edges of the stripes as observed for boundaries of F$\_{16}$CoPc structures on Ag(110)~\cite{Toader2010}. Both phases exhibit an oblique unit cell, where a coincidence of one of the lattice vectors of each phase is visible ($\vec{b}\_I=\vec{b}\_{II}$). Along this direction the molecules exhibit a long-range order, limited only by the boundaries of the covered region and step edges of the substrate. Remarkably, both phases exhibit only well-defined numbers of molecules in the second direction ($\vec{a}\_I, \vec{a}\_{II}$), which are always even in phase~I (e.g. 2 or 4) and odd in phase~II (e.g. 3 or 5). This was observed for all samples and very large regions of about $\SI{200}{nm} \times \SI{200}{nm}$.

Longer annealing seems to enlarge the regions, where a phase is continuous and solely present, as depicted in Fig.~\hyperref[fig:phases1]{\ref*{fig:phases1}(b)} and \hyperref[fig:phases1]{(c)}. These regions can show a \SI{60}{\degree} rotation with respect to each other due to the substrate symmetry. This can be seen easily in the Fast Fourier Transformation (FFT) images of those regions [Fig.~\hyperref[fig:phases1]{\ref*{fig:phases1}(d--f)}], where the pattern of the lattice spots of region II and II´ is rotated by \SI{60}{\degree}. In two adjacent regions (I and II), the lattices possess the same molecule--substrate relation, which can be seen in images~\hyperref[fig:phases1]{\ref*{fig:phases1}(d)} and \hyperref[fig:phases1]{(e)}, where the corresponding reciprocal lattice vectors ($\vec{a}\_I^*, \vec{a}\_{II}^*$) are equal.

Figure~\hyperref[fig:phases2]{\ref*{fig:phases2}(a)} shows two superimposed images of a covered (small tilted inset) and an uncovered (large) region of the same sample. Both images from the molecules and the HOPG were taken separately at different sites on the sample, less than \SI{100}{nm} away from each other. Therefore, we assume that the orientation of the HOPG did not change and a direct comparison of both lattices is possible. A comparison of the molecular lattices with the underlying lattice shows that the molecular lattice is incommensurate with respect to that of the HOPG. However, there is a commensurate match of the lattices in $\vec{b}\_I=\vec{b}\_{II}$ direction, as marked with small circles in the figure. No exact match in the other directions could be found, therefore it can be called a point-on-line coincidence~\cite{Hooks2001}. The lattice direction $\vec{a}\_{II}$ of phase~II is just slightly rotated by $\SI{3\pm 1}{\degree}$ with respect to the HOPG lattice. The common lattice direction $\vec{b}\_{I}=\vec{b}\_{II}$ exhibits an angle of $\SI{7.5 \pm 1}{\degree}$ with respect to the HOPG lattice. The lattice vectors of phase~I are $|\vec{b}\_I|\approx|\vec{a}\_I| = \SI{15\pm 1}{\angstrom}$ and for phase~II  $|\vec{b}\_{II}|=|\vec{b}\_I|=\SI{15\pm 1}{\angstrom}$  and  $|\vec{a}_{II}|= \SI{16.5\pm 1}{\angstrom}$. The angles between the vectors are $\alpha\_I = \SI{84\pm 3}{\degree}$ and $\alpha\_{II}=\SI{110\pm 3}{\degree}$. The angles differ by 3\,$^{\circ}$ depending on the particular sample and the accurate phase situation (mixed or undisturbed).
In the second and third monolayer, the same structures with their lattice constants and angles as in the first ML were observed (see figure~\hyperref[fig:phases1]{\ref*{fig:phases1}(a)} for a representative example).

As can be seen in figure~\hyperref[fig:phases2]{\ref*{fig:phases2}(b)}, the CoPc molecules of phase~I exhibit the same lobe orientation with respect to each other and an angle $\varphi\_3= \SI{62\pm 3}{\degree}$ between a lobe axis and the lattice (vector $\vec{b}\_I$). In phase~II the lobes exhibit two different orientations relative to $\vec{b}\_{II}$ with the angles  $\varphi\_1= \SI{29}{\degree}$ and $\varphi\_2= \SI{58}{\degree}$. The angle $\delta=\varphi\_2-\varphi_1$ between the lobes of two rows of molecules in phase~II is $\SI{29\pm 3}{\degree}$. Therefore, one could describe phase~I as a lattice with a single-molecular basis and phase~II with a two-molecular basis with the double lattice vector in $\vec{a}\_{II}$ direction. In figure~\hyperref[fig:phases2]{\ref*{fig:phases2}(b)} the extreme case, which was often found after post-annealing, is shown where phase~I consists only of one unit cell and, therefore, becomes the same as a dislocation line between two domains of phase~II [see also figure~\hyperref[fig:phases1]{\ref*{fig:phases1}(b)}].

\subsection{Model calculations}

\begin{figure}[!tb]
\centering
\includegraphics[width=0.95\textwidth]{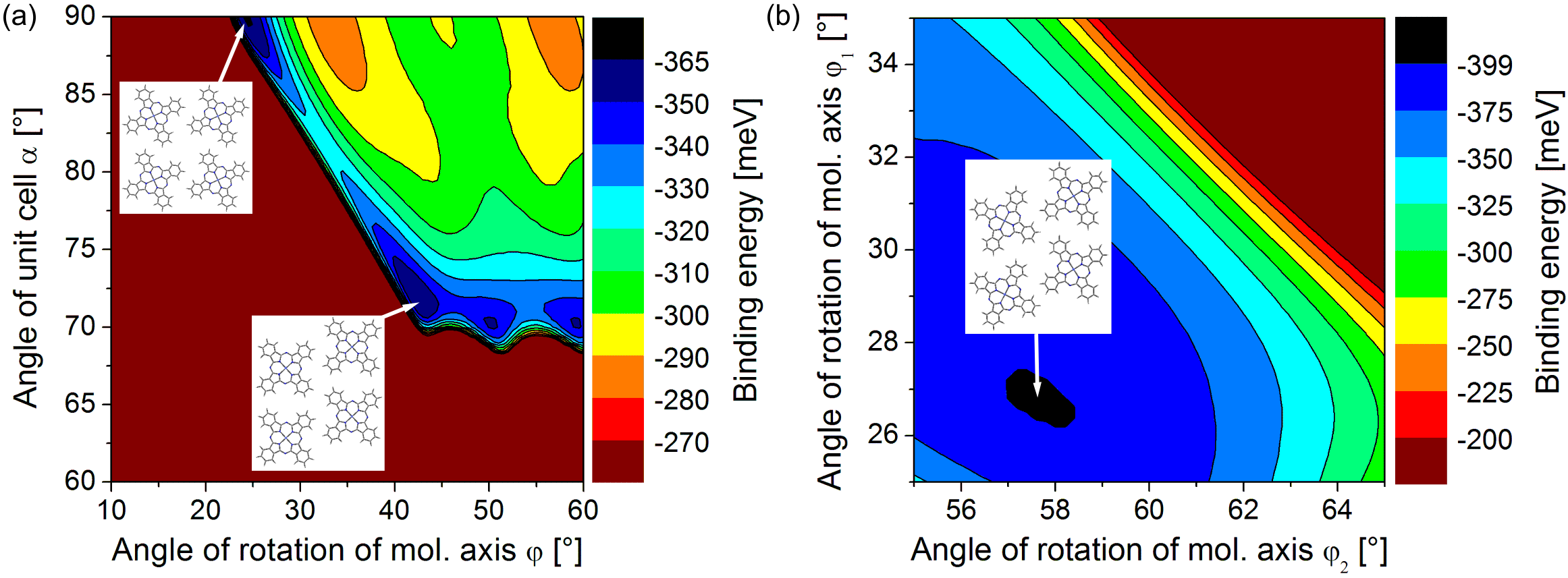}
\caption{Molecular mechanics calculations for the arrangement of CoPc.
(a) The potential energy surface for structures with the same orientation for all molecules at fixed intermolecular distances of $a = b = \SI{1.5}{nm}$ and the corresponding structures of the two lowest minima are shown.
(b) Potential energy surface for different orientations $\varphi\_1$ and $\varphi\_2$ of CoPc between molecular rows. The structure with the lowest energy is found for molecules rotated against each other at an unit cell angle of $\alpha = \SI{70}{\degree}$.}
\label{fig:UFF}
\end{figure}

Structures of CoPc molecules in an unsupported monolayer were calculated with UFF to validate the observed phases and identify the structure with the lowest energy. At fixed intermolecular distances of $|\vec{a}|=|\vec{b}|= \SI{1.5}{nm}$ and the same angle of rotation $\varphi = \varphi\_1 = \varphi\_2$ for all molecules [figure~\hyperref[fig:UFF]{\ref*{fig:UFF}(a)}], the structure with the lowest binding energy of \SI{-368}{meV} is found at $\varphi = \SI{24.5}{\degree}$ and an angle of the unit cell of $\alpha = \SI{90}{\degree}$. This structure is comparable with phase~I which was found with STM.  However, for the measured monolayer of CoPc on HOPG we can see that the interaction with the underlying substrate leads to a slight reduction of the unit cell angle to $\alpha\_I = \SI{84\pm 3}{\degree}$ which accounts for the adsorption on energetically more favorable graphite lattice positions. Also, the lobe orientation $\varphi$ is close to the measured value of about 28$^{\circ}$, whereby the deviation results from the smaller angle of the unit cell.

In the potential energy surface (Fig.~\ref{fig:UFF}) the structure with the second lowest minimum of \SI{-362}{meV} is found at $\alpha = \SI{72}{\degree}$ and $\varphi = \SI{47.5}{\degree}$. 
% The energetically lowest transition path from the energy minimum structure at $\alpha = \SI{90}{\degree}$ to that at $\alpha = \SI{72}{\degree}$ has a barrier height of \SI{40}{meV}. 
Furthermore, we found that, when molecules in a row have the same angle of rotation of the molecular axis $\varphi\_1$ but a different angle $\varphi\_2$ in the neighboring molecular row, structures with even lower binding energies occur [figure~\hyperref[fig:UFF]{\ref*{fig:UFF}(b)}]. The energetically most favorable molecular arrangement was found at an unit cell angle of \SI{70}{\degree} and the rotation angles of the lobes of the CoPc molecules of $\varphi\_1 = \SI{26.8}{\degree}$ and $\varphi\_2 = \SI{57.7}{\degree}$, as sketched in figure~\ref{fig:skizze}. The calculated binding energy for this structure is \SI{32}{meV} lower than that for the structure with the same rotation angle of all molecules and $\alpha = \SI{90}{\degree}$. Due to a mirror symmetry between the molecules and the underlying substrate, the calculations hold also true for the unit cell angle translated to $\alpha'=180^{\circ} - \alpha$. Therefore, the calculated \SI{70}{\degree} situation corresponds exactly to the measured phase~II ($\alpha\_{II} = \SI{100}{\degree} = \SI{180}{\degree}-\SI{70}{\degree}$) where the calculated angles $\varphi\_2$ and $\varphi\_1$ are close to the experimentally observed ones. The calculated angle $\delta = \varphi\_2 - \varphi\_1 = \SI{30.9}{\degree}$ is about \SI{2}{\degree} larger than the measured one, but still in good agreement with the experimentally observed phase~II. Since the substrate interaction is not taken into account, a deviation of the calculated angles for both lobe orientations from the measured ones is not surprising.

The above findings suggest that the intermolecular forces are the major driving factor for the structure formation, while the molecule--substrate interaction has only a small influence on the lattice parameters. This is similar to the findings on other phthalocyanines at HOPG, which also show a strong molecule-molecule interaction, that supersedes the molecule--substrate interaction in the forming of the molecular lattice on a surface~\cite{Gopakumar2006}. Structural investigations of SnPc~\cite{Walzer2001}, PdPc~\cite{Gopakumar2006} {and} FePc~\cite{Aahlund2007} on HOPG revealed chain building in the low coverage regime and the formation of two phases in the higher coverage regime. Again, the angles found are close to a quadratic and a hexagonal phase, but don't match exactly. A rotation of the single domains of \SI{30}{\degree} according to the HOPG hexagonal sublattice as well as a slight rotation of \SI{10}{\degree} of the molecular lattice with respect to the HOPG lattice were measured.

\section{Conclusions}

We investigated HOPG samples with ultrathin films of CoPc molecules. Below 1 ML the molecules show a preferred orientation along the substrates lattice and a short-range ordering in chains, while they are still very mobile on the surface. For the first monolayer we found two different structures. They are found also in the following layers because of direct molecular stacking. The structures exhibit a long-range order in the preferred direction, which leads to a point-on-line epitaxy with the underlying substrate lattice. The comparison between our experimental findings and molecular mechanics simulations yields a good agreement with a somewhat larger deviation for phase I which hints to a stronger molecule--substrate interaction as compared to phase~II.

\section*{Acknowledgements}

This paper has been funded by the Deutsche Forschungsgemeinschaft through the Research Unit FOR 1154 (Grants HA6037/1 and HI512/11) and the Graduate School GRK 1621.

\bibliographystyle{model1-num-names}
\bibliography{copc_hopg}

\end{document}